\documentclass[
aip,
rsi,
amsmath, amssymb,
reprint,
floatfix
]{revtex4-1}
\usepackage{graphicx}
\usepackage[hang,nooneline]{subfigure}
\usepackage{bm}
\usepackage[utf8]{inputenc}

\begin{document}

\preprint{}

\title{Minimizing magnetic fields for precision experiments}

\author{I.~Altarev}
\author{P.~Fierlinger}
\author{T.~Lins}
\author{M.~Marino}
\author{B.~Nießen}
\author{G.~Petzoldt}
\author{M.~Reisner}
\author{S. Stuiber}
\email{stefan.stuiber@ph.tum.de}
\author{M.~Sturm}
\author{J.~Taggart Singh}
\altaffiliation[now at ]{Michigan State University, 640 S. Shaw Lane, East Lansing, MI, USA}
\author{B.~Taubenheim}

\affiliation{Physikdepartment, Technische Universität München, D-85748 Garching, Germany}

\author{H. K. Rohrer}
\affiliation{Rohrer GmbH, D-80667 München, Germany}

\author{U. Schläpfer}
\affiliation{IMEDCO AG, CH-4614 Hägendorf, Switzerland}

\date{\today}

\begin{abstract}
An increasing number of measurements in fundamental and applied physics rely on magnetically shielded environments with sub nano-Tesla residual magnetic fields. 
State of the art magnetically shielded rooms (MSRs) consist of up to seven layers of high permeability materials in combination with highly conductive shields. 
Proper magnetic equilibration is crucial to obtain such low magnetic fields with small gradients in any MSR. 
Here we report on a scheme to magnetically equilibrate MSRs with a 10 times reduced duration of the magnetic equilibration sequence and a significantly lower magnetic field with improved homogeneity.
For the search of the neutron's electric dipole moment, our finding corresponds to a linear improvement in the systematic reach and a 40\% improvement of the statistical reach of the measurement.
However, this versatile procedure can improve the performance of any MSR for any application.
\end{abstract}


\maketitle 

\section{Introduction}
Magnetically shielded rooms (MSRs) are environments where external electromagnetic field distortions at low frequencies are strongly damped and the earth or ambient magnetic fields are strongly suppressed. MSRs are typically used in biomagnetic~\cite{trahms}, medical and fundamental physics applications~\cite{baker, tum_edm}. For these, both a strong suppression of time varying external fields of frequencies from mHz to kHz, as well as the resulting residual quasi static magnetic field ($<$~1~nT) and residual magnetic field gradients ($<$~1~nT/m) are important.
Shielding in the low frequency range of up to tens of Hz is usually achieved by using large amounts of highly permeable alloys (e.g. Permalloy). However, already at the 1~Hz regime the conductivity of the shielding material plays a significant role due to induced currents over large areas of the surface. For shielding of radio frequency (RF) disturbances, an additional layer of highly conducting material is added. The reference facility for magnetic shielding is the Berlin Magnetically Shielded Room 2 (BMSR-2) at the Physikalisch-Technische Bundesanstalt (PTB) Berlin~\cite{bmsr2}. With 7 layers of Mumetall\footnote{Mumetall is a brand name of Vacuumschmelze Hanau GmbH} and an additional aluminum layer for RF shielding it has a passive shielding factor (SF) of 75000 at 0.01~Hz. The new shield for the next-generation neutron electric dipole moment (nEDM) experiment at the Technische Universität München (TUM) reaches even higher SFs exceeding 1000000 at 1~mHz, without any active measures to compensate for low-frequency drifts~\cite{insert}.
Materials used for shielding have a high permeability and a non-zero remanence and therefore can be easily magnetized by arbitrary fields. Such fields are typically generated by (i) static external sources, (ii) internally applied fields for e.g.~in low-field NMR applications, (iii) magnetic fields emerging from adjacent shielding layers and (iv) magnetic distortions, e.g.~from construction materials of the shield itself. To achieve lowest possible residual fields and field gradients, the material has to be 'degaussed' repeatedly. 'Degaussing' is the commonly used term for the reduction of the remanent magnetization in magnetizable materials. It is usually performed by applying a sinusoidal current with decreasing amplitude to coils wound around the shielding material inducing magnetic flux in a closed loop of shielding material. The starting amplitude has to be large enough to reach saturation of the magnetization everywhere within the material. At the end of the cycle, the amlitude has to reach zero as exactly as possible. Any DC offset or noise can worsen the quality of the degaussing procedure. Degaussing of MSRs has been developed to an unprecedented quality~\cite{voigt, tum_msr}, with achieved residual fields of $<$~500~pT and gradients of $<$~0.2~nT/m over the center cubic meter volume inside walk-in MSRs. These small fields can be generated in a highly reproducible manner to within 10’s of pT, independent of the initial conditions.

The relationship between residual field and shielding has also been well investigated~\cite{tum_msr}. The shielding factor of an MSR is given as the frequency dependent attenuation of external disturbances
\begin{equation}
SF(\omega)  = \frac{A_o \times \sin(\omega t)} {A_i \times \sin(\omega t + \phi)}.
\end{equation}
$A_o$ denotes the field magnitude at a position measured before the presence of the MSR, $A_i$ the magnitude at the same position with the MSR, $\omega$ is the frequency of the external disturbance and $\phi$ is a phase shift between the excitation field and the measured field inside the MSR. Assuming a \emph{sufficient} amount of shielding material, in the static case ($\omega=0$) the field within the MSR is dominated rather by the residual field from the shielding material, and \emph{not} by the effect of external fields.
Ultra-low residual fields are achieved by effectively offering the magnetic field a path of minimal magnetic resistance around the volume to be shielded. By minimizing the potential energy of the magnetic domains within the shielding material in their surrounding field an equilibrium state is formed, which is stable in time for a static environment. Therefore degaussing does not per se remove the magnetic field as the name suggests, but rather optimizes the magnetic field in the material with respect to the external condition. For these reasons, we prefer the usage of 'magnetic equilibration' instead of degaussing to refer to this process (naturally, spots of strongly magnetized material still have to be degaussed to reduce their magnetization).
One of the procedures to minimize residual magnetic fields inside MSRs is described in Ref.~\onlinecite{voigt}. Here, the coil setup for equilibration consists of 12 coils for each layer of the MSR with one along each edge of the cube. Four of these coils are connected in series to form a closed loop around one spatial direction (see also Fig~\ref{fig:coils}a). The three spatial directions are then degaussed sequentially. A sinusoidal current with a peak amplitude of 21$\times$5~Ampere$\times$turns and a frequency of 10~Hz is applied. The amplitude decreases over 2000 periods (‘cycles’) to a value as close to zero as technically achievable. Equilibration therefore takes 600~s for each layer. With this procedure residual fields below 1.5~nT and gradients smaller than 2~nT$/$m have been achieved.
In Ref.~\onlinecite{thiel} one of the main criteria for a reproducible procedure is the step size of the amplitude decrease. The difference in amplitude between two subsequent maxima of the sinusoidal current should be as small as possible as this delta corresponds to an error in the residual field. Therefore it is suggested to use several thousand cycles.\\
\indent In this work we describe a new procedure for magnetic equilibration, which is capable of producing extremely small residual fields, comparable to the best ever achieved. However, the result is achieved within a much shorter time constant of only 30-50~s per layer, with simpler wiring of equilibration coils and with simpler equipment. In the previous procedure with a duration of about 900~s, residual fields of $<$~0.7~nT and field gradients of 0.3~nT/m were achieved in the MSR at TU München~\cite{tum_msr}. The new configuration can produce the same or better values in only 150~s. Here, both layers of the MSR are equilibrated in a defined sequence.
For MSRs with only one layer of magnetizable material, such a sequence can even be shorter, on the order of only 10~s with a resulting residual field of $<$10~nT over a cubic meter inside, making single layer shielding a highly competitive technical solution at low cost.
Our finding has consequences for several applications, including in fundamental physics where the duty cycle of experiments is affected by frequent magnetic equilibration sequences (e.g.~required due to frequent reversal of applied NMR fields inside). Another example would be applications in biomagnetism, where the door to an MSR may be opened frequently. 

\section{Apparatus}
\begin{figure}
 \includegraphics[width=0.3\textwidth]{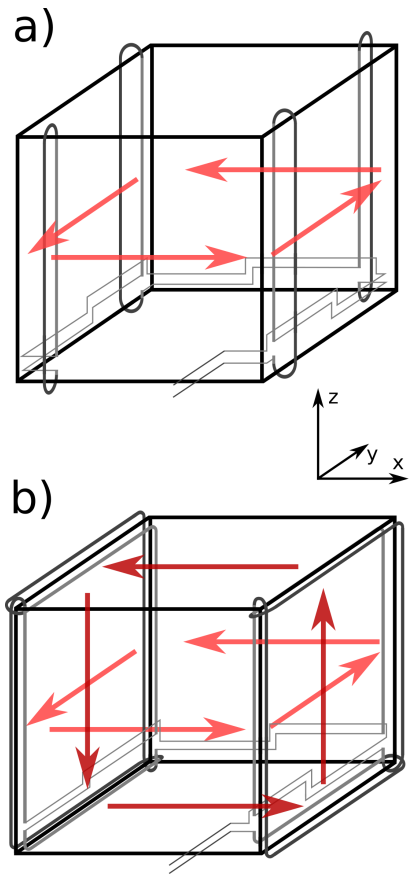}
 \caption{The setup of the equilibration coils for the old configuration, shown in a), and the new configuration in b). In a) the 4 coils to create a magnetic flux around the Z direction and their connections are shown. The new configuration in b) has the old Y and Z direction connected in a L shape so that a magnetic flux is created in both directions simultaneously.}
 \label{fig:coils}
\end{figure}
\begin{figure}
 \includegraphics[width=0.35\textwidth]{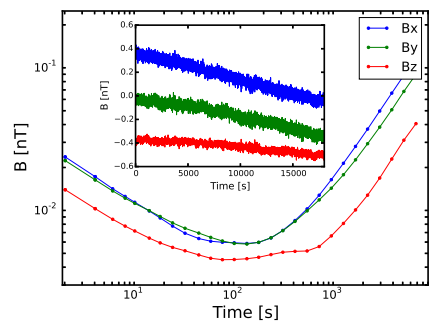}
 \caption{Allan deviation~\cite{allan} calculated for data recorded with the fluxgate inside the MSR for about 20000~s. The inset shows the corresponding time data, shifted to 0 at $t~=~0$ on the left axis. Bx and Bz have then been offset by 0.4 nT and -0.4 nT respectively for clarity. The smooth change of the magnetic field over time correlates with a smooth, monotonic drop of $0.2^\circ$~C in the environmental temperature. To attribute the correlation of field and temperature to a change of permeability is however more complex.}
 \label{fig:fg}
\end{figure}
The MSR used for this measurement and also the previous configuration of equilibration coils are described in Ref.~\onlinecite{tum_msr} (see also Fig \ref{fig:coils}a). The two layers of the MSR are equilibrated individually, first the inner layer, then the outer one and after that the inner one again. For each layer, we start with the X direction followed by the Z direction and then the Y direction (see also Fig \ref{fig:coils}). It was found that equilibration of one direction is not sufficient: subsequent equilibrations in the second and third direction each improved the result. Note the door is in the XZ-plane at negative y coordinate.
Each equilibration sequence starts with an amplitude that saturates the magnetizable material, in the case of the inner layer that is $9 \times 7$ Ampere turns, and then linearly decreases over 1000~cycles to below noise level. A frequency of 10~Hz is used. The waveform for equilibration is generated in software on a PC, converted to an analog voltage by a 16~bit digital-analog converter (DAC) with a sampling frequency of 20~kHz. This voltage is passed through an analog voltage divider to use the full range of the DAC and a 100~Hz lowpass filter to smooth out DAC steps. It is converted to a current in a power amplifier\footnote{Custom power amplifier model 'Rohrer PA 2088B' provided by Rohrer GmbH Munich} and passed through a transformer to remove any DC offset before being distributed to the respective coil of the MSR by an automated relay switch. In this way magnetic equilibration can be done completely automated, in a process that takes about 15~min.

Field maps have been measured using a three-axis fluxgate sensor\footnote{Bartington Mag03-IEHV70}. The sensor heads at the end of flying leads are mounted in a nonmagnetic structure so that they form a right handed coordinate system. Their signal is read out with a 16-bit analog digital converter (ADC). Each data point in the maps is an average over a measuring time of 1~s, where data has been recorded with a sampling frequency of 1~kHz. Several measurements taken at the same point deviate less than 0.1~nT from each other. To correct for drifts of fluxgate offsets, a calibration is done before, during and after each field map.
The sensor is calibrated in the center of the room where the fields are smallest, to keep the influence of alignment errors as small as possible. To determine the offset of a sensor the field is measured at a position and then again at the same position with the sensor rotated by $180^{\circ}$ to measure in the opposite direction. Assuming a linear drift between calibrations the offsets drift rates were $<$~10~pT/min. All maps were corrected for these offset drifts.

Figure \ref{fig:fg} shows the performance of the fluxgate magnetometer using an Allan variance plot. To obtain the optimal performance, an integration time of about 100~s would be required. For shorter times, noise will have an increasing effect on the measured signals. For longer times the measurement will be dominated by drifts that could come from changing offsets or drifts of the readout electronics. To keep the measurement time per point reasonably short, an integration time of 1~s was chosen. Here the Allan deviation is already below 30~pT. With this and repeated offset calibration an overall accuracy of $\sim$~0.3~nT can be reached, so that low-noise fluxgate sensors are well suitable for this kind of measurements.

After magnetic equilibration in the earth field without active stabilization, field maps were recorded for the respective configuration. Each map consists of 108 points on a 6~by~6~grid in 3 different XY planes (for coordinate system see Fig.~\ref{fig:coils}) in the center of the MSR, 35~cm below and 35~cm above it. The distance between the grid points along each axis is 10~cm. The sensor was moved by hand from one point to the next with an estimated accuracy of about 0.5~cm and a tilt-precision of better than few degrees. The distance of the individual sensors in the mount is 2~cm, so the overall position accuracy of the measured points is about 2~cm.

\section{New coil configuration and measurements}
\begin{figure}
 \includegraphics[width=0.5\textwidth]{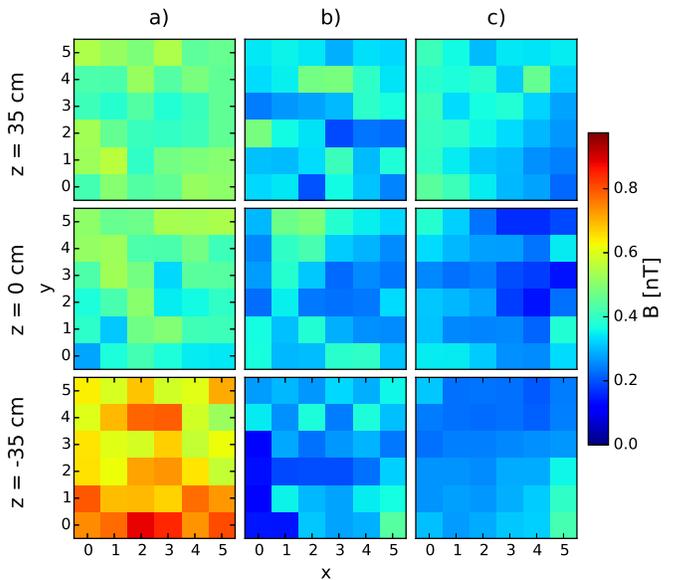}
 \caption{Maps of the magnitude of the magnetic flux density inside the MSR after magnetic equilibration in three XY planes at different heights. The points are 10~cm apart in each direction. The columns show: the previous coil configuration (a), and the L shaped configuration ( (b) and (c) ), where the later were taken with several days between the two measurements.} 
 \label{fig:maps}
\end{figure}
The equilibration scheme we describe here uses a different coil configuration: two coils sets of the previous configuration are reconnected so that the coils form an ‘L-shaped’ coil pattern along two edges of the MSR. Four of these Ls are again connected in series (see Fig. \ref{fig:coils} b). All connections between the coils are routed in a way that the fields produced by the connecting wires cancel. In this way, two spatial directions are equilibrated at the same time. The results show that the third direction can be omitted. Only the inner layer of the MSR was modified in this way. The equilibration sequence starts with the inner layer, followed by the unmodified outer layer and inner layer in sequence, using 50~s for the inner layer. This time can be further reduced, but below 30~s a notable decrease in the remanent field quality is observed. 
Figure \ref{fig:maps} shows the recorded field maps. In column a) the residual field for the old configuration is shown. The maximum value is about 0.8 nT in the lower plane and near the door of the MSR. In the upper planes the field is about 0.5~nT. The higher values close to the door are likely caused by stray fields of the last equilibration step. It was previously~\cite{tum_msr} found that equilibrating the direction not including the door yielded the best result. However there stray field again magnetize the overlaps. Columns b) and c) show the residual field for the new configuration, measured twice with several days in between to show reproducibility. Here the residual field is slightly smaller, on the order of 0.5~nT or below. Also the prominent feature in the lower plane is not visible anymore. With only one equilibration cycle a re-magnetization of the adjacent wall is avoided. These are values that are achieved in typical daily operation, where non-magnetic furniture and tools, as well as people are inside the MSR.

\begin{figure}
 \includegraphics[width=0.3\textwidth]{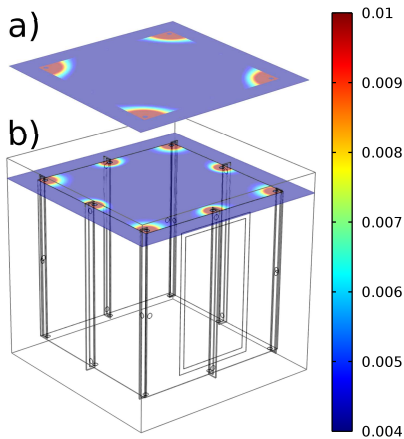}
 \caption{Simulations of the magnetization in the adjacent face that is not part of the equilibration loop. The geometry of the room and placement of the coils is the same for both simulations and only shown in b). In a) a current is supplied to the four coils at the edges, in b) all eight coils carry a current. The color scale is the same for both plots and in arbitrary units. Both cases illustrate that the currents create a significant magnetization in the adjacent face.}
 \label{fig:simcoils}
\end{figure}
To illustrate problems with the placement of equilibration coil a finite-elemente-method simulation with COMSOL is shown in Fig.~\ref{fig:simcoils}. In a) a current is applied to the four coils at the edges of the MSR, creating a flux loop around the four side walls of the MSR (around the Z direction). In Fig.~\ref{fig:simcoils}b) eight coils distributed equally along the path of flux are supplied with half the current used for a). Each coil has to penetrate the top wall of the MSR to form a loop with shielding material enclosed. The simulation shows the magnetization inside the top wall created by the current through the coils. Since this face is not included in the equilibration loop, its magnetization is not properly removed by the equilibration process. This dominates the error caused by a possible DC offset at the level of precision reached in our setup. Here a big number of equilibration cycles (as mentioned in Ref.~\onlinecite{thiel}) reduce this error. Additionally, when small volumes in the center of the MSR are considered, the error has a small influence on the residual field due to the symmetry of the MSR. However for extended-size volumes these contributions become more significant. Even when currents are better distributed and have a smaller amplitude the adjacent face retains a magnetization. By equilibrating all faces at once this problem does not occur.

\section{Conclusion}
The new scheme for equilibration coils presented here produces static remanent fields below 0.5~nT inside a two-layer MSR with less contributions from magnetizable features at the walls. This result is on the same order of magnitude as that from the previous equilibration setup, but achieved with significantly shorter cycle times. For the two-layered room at TUM, the time will be reduced from 900~s to 150~s when both layers are changed. In an MSR with more layers, the reduction can be even larger. For the outer layers perfect equilibration is not necessary and the time can be shorter than the 50~s used here. If for example the outer three layers of a 5 layer MSR are equilibrated for 20~s, and the inner two for 50~s, the whole procedure would take only 160~s instead of 1500~s. In the nEDM experiment at TUM the shield has to be equilibrated for example every 10~min. With the reduced time for the equilibration, the duty cycle of the experiment is effectively increased by a factor of 2, resulting in an improved sensitivity of the measurement by about 40\%.
In a single layer shield, where only remanent fields of tens of nT are targeted, equilibration can be done within about 10~s. Additionally, this setup only requires one set of coils per layer. This significantly reduces the amount of high current relays needed for automated switching between the equilibration coils. 
Further, the simultaneous equilibration of all faces avoids accidental magnetization in adjacent faces due to stray fields.

\begin{acknowledgments}
We acknowledge the support at the FRM-II research reactor in Garching. This work was supported by DFG Priority Program SPP 1491 and the DFG Cluster of Excellence 'Origin and Structure of the Universe'.
\end{acknowledgments}


%

\end{document}